\documentclass[pra,11pt,tightenlines,nofootinbib,superscriptaddress]{revtex4}

\usepackage{graphicx}
\usepackage{dcolumn}
\usepackage{bm}

\begin{document}

\title{Modification of radiation pressure due to cooperative scattering of light}

\author{Ph.W. Courteille}
\affiliation{Physikalisches Institut, Universit\"at T\"ubingen, 72076 T\"ubingen, Germany}
\affiliation{Instituto de F\'isica de S\~ao Carlos, Universidade de S\~ao Paulo, 13560-970 S\~ao Carlos, SP, Brazil}
\affiliation{Institut Non Lin\'eaire de Nice, CNRS, Universit\'e de Nice Sophia-Antipolis, 06560 Valbonne, France}

\author{S. Bux}
\affiliation{Physikalisches Institut, Universit\"at T\"ubingen, 72076 T\"ubingen, Germany}
\affiliation{Institut Non Lin\'eaire de Nice, CNRS, Universit\'e de Nice Sophia-Antipolis, 06560 Valbonne, France}

\author{E. Lucioni}
\affiliation{Institut Non Lin\'eaire de Nice, CNRS, Universit\'e de Nice Sophia-Antipolis, 06560 Valbonne, France}
\affiliation{Dipartimento di Fisica, Universit\`a Degli Studi di Milano, Via Celoria 16, 20133 Milano, Italy}

\author{K. Lauber}
\affiliation{Physikalisches Institut, Universit\"at T\"ubingen, 72076 T\"ubingen, Germany}

\author{T. Bienaim\'e}
\affiliation{Institut Non Lin\'eaire de Nice, CNRS, Universit\'e de Nice Sophia-Antipolis, 06560 Valbonne, France}

\author{R. Kaiser}
\affiliation{Institut Non Lin\'eaire de Nice, CNRS, Universit\'e de Nice Sophia-Antipolis, 06560 Valbonne, France}

\author{N. Piovella}
\affiliation{Dipartimento di Fisica, Universit\`a Degli Studi di Milano, Via Celoria 16, 20133 Milano, Italy}

\date{March 2010}

\begin{abstract}
Cooperative spontaneous emission of a single photon from a cloud of $N$ atoms modifies substantially the radiation pressure exerted by a far-detuned laser beam exciting the atoms. On one hand, the force induced by photon absorption depends on the collective decay rate of the excited atomic state. On the other hand, directional spontaneous emission counteracts the recoil induced by the absorption. We derive an analytical expression for the radiation
pressure in steady-state. For a smooth extended atomic distribution we show that the radiation pressure depends on the atom number via cooperative scattering and that, for certain atom numbers, it can be suppressed or enhanced. Cooperative scattering of light by extended atomic clouds can become important in the presence of quasi-resonant light and could be addressed in many cold atoms experiments.
\end{abstract}

\maketitle

\section{Introduction}

Coherent radiation by large collections of small particles has been studied extensively with applications in antennas, atmospheric scattering or plasma diagnostics. The
behavior of those macroscopic scattering devices is ruled by the interference of the fields scattered by many microscopic elements.
In atomic physics, matter-light interactions have been studied extensively
in the past decades, ranging from spectroscopy to laser manipulation of the atomic internal and external degrees
of freedom. Intriguing properties can be obtained
when many atoms are interacting with photons, as studied in the seminal work by Dicke \cite{Dicke54}
and more recently by cold samples of bosonic atoms in a trap \cite{You96}.
Other complex quantum features are obtained when the quantum properties of light and/or of atoms are
taken into account, as is done e.g. in studies of quantum cryptography and quantum computation \cite{Cirac2008}.
Often a quantum description of two level atoms and a quantum description
of the light field in terms of photon operator is very convenient to describe the atom-light interactions. However
one should not forget that many features can well be described with classical models for the atoms with a
polarizability  and for the light field. Many features of Dicke superradiance can thus be explained using classical
antenna theory. As atoms appear to be excellent systems to study any possible deviation from classical many
body features, it is a general interest to understand and to control cooperative effects, even at a classical level,
in a cloud of atoms. New intriguing effects can arise, when fluctuations due to the coupling with the vacuum modes can
no longer be described by a classical field approach and the atoms can also become entangled during the cooperative scattering. Such cooperative scattering with atomic ensembles can appear in a number of experimental situations in free space \cite{Javanainen94} or in cavities \cite{Raimond82}.

\begin{figure}[ht]
        \centerline{\scalebox{.7}{\includegraphics{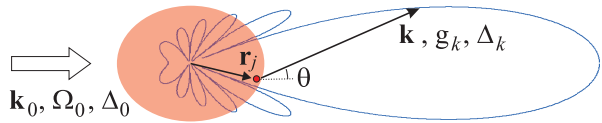}}}
        \caption{(Color online) Scheme for collective scattering of a single photon from a pump laser beam $\mathbf{k}_0$ by a spatial distribution of atoms.}
        \label{Fig1}
    \end{figure}

The problem of collective spontaneous emission has recently received growing interest with the study of single photon superradiance from $N$ two-level atoms prepared by the absorption of one photon of wave vector $\mathbf{k}_0=k_0\mathbf{e}_z$ \cite{Scully06,Scully09,Eberly06,Glauber08}. It has been shown that the photon is spontaneously emitted in the same direction of the incident photon with a cooperative decay rate proportional to $N$ and inversely proportional to the size of the atomic cloud \cite{Svidzinsky08}. These studies considered the decay of atoms prepared in the `timed Dicke state':
\begin{equation}\label{EqDicke}
    |+\rangle_{\mathbf{k}_0}=\frac{1}{\sqrt{N}}\sum_{j=1}^N\,e^{i\mathbf{k}_0\cdot\mathbf{r}_j}|g_1,g_2,\dots,e_j,\dots,g_N\rangle,
\end{equation}
where $|g_1,g_2,\dots,e_j,\dots,g_N\rangle$ is a Fock state in which the atom $j$ is prepared in the excited state $e$ and all the other atoms being in the ground state $g$, and $\mathbf{r}_j$ is the position of the atom $j$. In these papers, only resonant photons have been considered and their reabsorption by other atoms has been neglected. As a situation closer to an experimental realization, we consider a cold atomic cloud with a Gaussian density distribution irradiated by a plane wave laser beam (see Fig.~\ref{Fig1}). If the laser beam is detuned far enough from the atomic resonance to avoid multi-atom absorption, a timed Dicke state is generated. The large detuning also ensures that the dominant fraction of photons is elastically scattered.

In this paper, going beyond the mere consideration of the light emission by an atomic ensemble, we will revisit the phenomenon of radiation pressure in a regime where scattering is dominated by cooperative effects. Mechanical effects combined to cooperative emission have been studied for atomic ensembles in cavities \cite{Vuletic00} or in optical lattices \cite{Deutsch1994}. However, they also occur in free space \cite{Bonifacio94} and, as we will show in this paper, can modify not only the emission properties but also the absorption process. Indeed, the absorption properties are affected by superradiant enhancement of the spontaneous decay rate. On the other hand, cooperative scattering bundles the scattered light into forward direction, which leads to a suppression of radiation pressure. This suppression is only limited by the inhomogeneity of the atomic cloud, which occupies a finite volume. For volumes larger than the Dicke limit, $|\mathbf{r_j}|\gg\lambda$, the net effect of cooperativity is a suppression of radiation pressure far below the value expected for independent scattering.

We proceed as follows. We solve the Schr\"odinger equation for an ensemble of $N$ two-level atoms  with the resonance frequency $\omega_a$ driven by a far-detuned laser field with the frequency $\omega_0\equiv\omega_a+\Delta_0$ and the wave vector $\mathbf{k}_0$. For a laser field sufficiently detuned from resonance so that the average number of excited atoms in the cloud is less than unity, $\sqrt{N}\Omega_0<\Delta_0$ (where $\Omega_0$ is the Rabi frequency of the laser field), we assume that only one among $N$ atoms is excited by absorption of a single laser photon, which is then spontaneously emitted into vacuum modes with frequencies $\omega_k\equiv\omega_a+\Delta_k$ and wave vectors $\mathbf{k}$. Within the Markov approximation (valid for $t\gg R/c$, where $R$ is the dimension of the cloud \cite{Svidzinsky08}) we obtain expressions for the time-evolution of the state amplitudes allowing us to derive a generalized expression for the steady-state radiation pressure in the presence of cooperative effects.

In the limit where the atomic cloud can be described by a smooth spherical Gaussian density distribution, an analytical calculation of the structure factors allows us to derive an explicit expression for the radiation pressure as a function of volume, shape, atom numbers, and pump laser detuning. In particular, we identify two regimes distinct by different scalings of the radiation pressure with atom numbers.

\section{Time evolution of the system}

The atom-field interaction Hamiltonian is in the rotating-wave approximation (RWA) \cite{Note01}
\begin{eqnarray}\label{EqHamiltonian}
    \hat{H}  =\hbar\sum_{j=1}^N\left[\frac{\Omega_0}{2}\hat{\sigma}_j e^{i\Delta_0t-i\mathbf{k}_0\cdot\mathbf{r}_j}+\textrm{h.c.}\right]
     +\hbar\sum_{j=1}^N\sum_{\mathbf{k}}\left[g_k\hat{\sigma}_j\hat{a}_{\mathbf{k}}^\dagger e^{i\Delta_kt-i\mathbf{k}\cdot\mathbf{r}_j}
        +\textrm{h.c.}\right]~.
\end{eqnarray}
Here, $\Omega_0$ is the Rabi frequency of the interaction between an atom and the pump mode (which is treated as a classical field), $\hat{\sigma}_j$ is the lowering operator for atom $j$, $\hat{a}_{\mathbf{k}}$ is the photon annihilation operator, and $g_k =d\sqrt{\omega/(\hbar\epsilon_0V_{ph})}$ describes the coupling between the atom and the vacuum modes with volume $V_{ph}$. The $j$-th atom has lower and upper states denoted by $|g_j\rangle$  and $|e_j\rangle$, respectively. With this approach we treat the atoms as simple two-level systems. We also assume that all atoms are driven by the unperturbed incident laser beam, thus neglecting dephasing by atoms along the laser path or by near field effects, which could arise for large spatial densities. Calling $|0\rangle_a=|g_1,..,g_N\rangle$ the atomic ground state and $|j\rangle_a=|g_1,..,e_j,..,g_N\rangle$ the state where only the atom $j$ is excited, we assume that the total state of the system has the following form:
\begin{eqnarray}\label{EqAnsatz}
    |\Psi(t)\rangle =\alpha(t)|0\rangle_a|0\rangle_{\mathbf{k}}
        + \sum_{j=1}^N~\beta_j(t)|j\rangle_a|0\rangle_{\mathbf{k}}+\sum_{\mathbf{k}}\gamma_{\mathbf{k}}(t)|0\rangle_a|1\rangle_{\mathbf{k}}~.
\end{eqnarray}
The above expression assumes that only states with at most one atomic excitation contribute to the effects described in this paper.

The time evolution of the amplitudes is obtained by inserting the Hamiltonian (\ref{EqHamiltonian}) and the ansatz (\ref{EqAnsatz}) into the Schr\"odinger equation, $\partial_t|\Psi(t)\rangle =-(i/\hbar)\hat{H}|\Psi(t)\rangle$:
\begin{eqnarray}
    \dot{\alpha}(t)  =-\frac{i}{2}\Omega_0e^{i\Delta_0t}\sum_{j=1}^N\beta_j(t)e^{-i\mathbf{k}_0\cdot\mathbf{r}_j}~,\label{EqAmplitude1}\\
    \dot{\beta}_j(t) =-\frac{i}{2}\Omega_0\alpha(t)e^{-i\Delta_0t+i\mathbf{k}_0\cdot\mathbf{r}_j}
        -i\sum_{\mathbf{k}}g_k\gamma_{\mathbf{k}}(t)e^{-i\Delta_kt+i\mathbf{k}\cdot \mathbf{r}_j}~,\label{EqAmplitude2}\\
    \dot{\gamma}_{\mathbf{k}}(t) =-ig_ke^{i\Delta_kt}\sum_{j=1}^N\beta_j(t)e^{-i\mathbf{k}\cdot\mathbf{r}_j}~.\label{EqAmplitude3}
\end{eqnarray}
Integrating Eq.~(\ref{EqAmplitude3}) over time and substituting $\gamma_{\mathbf{k}}(t)$ in Eq.~(\ref{EqAmplitude2}) we obtain
N coupled equations:
\begin{eqnarray}\label{EqEvolution}
    \dot{\beta}_j(t)= - \frac{i}{2}\Omega_0\alpha(t)e^{-i\Delta_0t+i\mathbf{k}_0\cdot\mathbf{r}_j}
     -\sum_{\mathbf{k}}g_k^2\sum_{m=1}^Ne^{i\mathbf{k}\cdot(\mathbf{r}_j-\mathbf{r}_m)}\int_0^{t}e^{-i\Delta_k(t-t')}\beta_m(t')dt'~.
\end{eqnarray}
We have numerically checked that due to the presence of the driving term, the solution quickly evolves toward a ``driven timed Dicke state" \cite{Scully06}, characterized by
\begin{equation}\label{EqTimed}
    \beta_j(t)=\frac{\beta(t)}{\sqrt{N}}e^{-i\Delta_0t+i\mathbf{k}_0\cdot\mathbf{r}_j}.
\end{equation}
Once inserted the ansatz (\ref{EqTimed}) in Eqs.~(\ref{EqAmplitude1}),(\ref{EqEvolution}) and going to continuous momentum space via $\sum_{\mathbf{k}}\rightarrow
V_{ph}(2\pi)^{-3}\int d\mathbf{k}$, we get
\begin{equation}
    \dot\alpha(t)= -\frac{i}{2}\sqrt{N}\Omega_0\beta(t)~,\label{EqMotion1}
\end{equation}
\begin{equation}
    \dot\beta(t) = -\frac{i}{2}\sqrt{N}\Omega_0\alpha(t)+i\Delta_0\beta(t)-\frac{V_{ph}N}{(2\pi)^3}\int d\mathbf{k}~g_k^2|S_N(\mathbf{k})|^2
    \times \int_0^te^{-i\left(\omega_k-\omega_0\right)t'}\beta(t-t')dt'~,\label{EqMotion2}
\end{equation}
where $S_N(\mathbf{k})\equiv\frac{1}{N}\sum_{j=1}^Ne^{-i(\mathbf{k}-\mathbf{k}_0)\cdot\mathbf{r}_j}$ is the structure factor of the atomic cloud. We can now apply the Markov approximation and write
\begin{eqnarray}\label{EqMarkov}
    \int_0^te^{-i\left(\omega_k-\omega_0\right)t'}\beta(t-t')dt' & \approx \frac{\pi}{c}\delta(k-k_0)\beta(t).
\end{eqnarray}
Then, Eq.~(\ref{EqMotion2}) becomes
\begin{equation}
    \dot{\beta}(t)=-\frac{i}{2}\sqrt{N}\Omega_0\alpha(t)+\left(i\Delta _0-\frac{1}{2}\Gamma Ns_N\right)\beta(t)~,
\end{equation}
using the definition of $\Gamma\equiv(V_{ph}/\pi c)k_0^2g_{k_0}^2$. Here, using spherical coordinates, we introduced the quantity $s_N$:
\begin{equation}\label{EqsN}
    s_N = \frac{1}{4\pi}\int_0^{2\pi}d\phi\int_0^{\pi}d\theta\sin\theta\left|S_N(k_0,\theta,\phi)\right|^2~.
\end{equation}
In steady state and neglecting saturation, $\sqrt{N}\Omega_0\ll\Delta_0$, we may assume $\alpha(t)\approx 1$ and find the
\begin{eqnarray}\label{EqSteadybeta}
    \beta^{st}  \approx\frac{\sqrt{N}\Omega_0}{2\Delta_0+iN\Gamma s_N}~.
\end{eqnarray}

\section{Forces in the Markov approximation}

The two terms in the Hamiltonian~(\ref{EqHamiltonian}) yield two different contributions to radiation pressure force:
\begin{equation}
    \mathbf{\hat{F}}_{aj}+\mathbf{\hat{F}}_{ej}=-\nabla_{\mathbf{r}_j}\hat{H}~.
\end{equation}
We will be interested in the average absorption force, $\mathbf{F}_a=\frac{1}{N}\sum_j\langle\mathbf{\hat F}_{aj}\rangle$, and emission force, $\mathbf{F}_e=\frac{1}{N}\sum_j\langle\mathbf{\hat F}_{ej}\rangle$, acting on the center of mass of the whole cloud, $\mathbf{F}_a+\mathbf{F}_e=m{\mathbf{a}}_{CM}$, where $\mathbf{a}_{CM}$ is the center-of-mass acceleration and $m$ the mass of one atom.

The first term, $\mathbf{\hat{F}}_{aj}=\frac{i}{2}\hbar\mathbf{k}_0\Omega_0[\hat{\sigma}_je^{i\Delta_0t-i\mathbf{k}_0\cdot\mathbf{r}_j}-\textrm{h.c.}]$ results from the recoil received upon absorption of a photon from the pump laser and has an expectation value on the timed Dicke state given by:
\begin{equation}\label{EqAbsorptionforce}
   \mathbf{F}_a=\langle\mathbf{\hat{F}}_{aj}\rangle=\frac{\hbar\mathbf{k_0}\Omega_0}{\sqrt{N}}\text{Im}\left[\alpha(t)\beta^*(t)\right]~.
\end{equation}
The second contribution, $\mathbf{\hat{F}}_{ej}=i\sum_{\mathbf{k}}\hbar\mathbf{k}g_k[\hat{\sigma}_j\hat{a}_{\mathbf{k}}^{\dagger}e^{i\Delta_kt
-i\mathbf{k}\cdot\mathbf{r}_j}-\textrm{h.c.}]$, results from the emission of a photon into any direction $\mathbf{k}$. The expectation value on the general state (\ref{EqAnsatz}) is:
\begin{equation}\label{EqStructuregauss}
    \langle\mathbf{\hat{F}}_{ej}\rangle=i\sum_{\mathbf{k}}\hbar\mathbf{k}g_k\left[\beta_j(t)\gamma_{\mathbf{k}}^\ast(t)e^{i\Delta_kt-i\mathbf{k}
        \cdot\mathbf{r}_j}-\textrm{c.c.}\right]~.
\end{equation}
Substituting the time integral of $\gamma_{\mathbf{k}}(t)$ from Eq.~(\ref{EqAmplitude3}) and inserting the timed Dicke state from Eq.~(\ref{EqTimed}) we obtain the average emission force:
\begin{eqnarray}
   \mathbf{F}_e = & -\sum_{\mathbf{k}}\hbar\mathbf{k}g_k^2|S_N(\mathbf{k})|^2
    \times \left[\beta(t)\int_0^te^{i\left(\omega_k-\omega_0\right)t'}\beta^{\ast}(t-t')dt'+\textrm{c.c.}\right]
\end{eqnarray}
In the Markov approximation and going to continuous momentum space,
\begin{eqnarray}
    \mathbf{F}_e & = -|\beta(t)|^2\frac{V_{ph}}{4\pi^2c}\int d\mathbf{k}(\hbar\mathbf{k})g_{k}^2|S_N(\mathbf{k}))|^2\delta(k-k_0)
\end{eqnarray}
Defining
$ f_N = \frac{1}{4\pi}\int_0^{2\pi} d\phi\int_0^{\pi}d\theta\sin\theta\cos\theta\left|S_N(k_0,\theta,\phi)\right|^2$,
we find
\begin{equation}\label{EqEmissionforce}
    \mathbf{F}_e=-\hbar\mathbf{k}_0\Gamma\left|\beta(t)\right|^2 f_N~.
\end{equation}
Finally, using Eq.~(\ref{EqSteadybeta}) and $\alpha\approx 1$ in Eqs.~(\ref{EqAbsorptionforce}) and (\ref{EqEmissionforce}), the steady-state value of the total radiation pressure force on the center of mass of the atomic cloud is
\begin{eqnarray}\label{EqCooperativeforce}
    F_c & \equiv F_a+F_e = \hbar k_0\Gamma\frac{N\Omega_0^2}{4\Delta_0^2+N^2\Gamma^2s_N^2}~(s_N-f_N)~.
\end{eqnarray}
Both forces, $\mathbf{F}_a$ and $\mathbf{F}_e$, are directed along the direction of the incident laser beam.

This is the main result of the paper. We identify contributions from cooperative absorption and cooperative emission. The common prefactor can be obtained from the standard single-atom radiation pressure force $F_1 = \hbar k_0 \Gamma\frac{\Omega_0^2}{4\Delta_0^2+\Gamma^2}$,
by substituting in the fraction the natural linewidth by the collective linewidth, $\Gamma\rightarrow N\Gamma s_N$, and the Rabi frequency by the collective Rabi frequency, $\Omega_0\rightarrow\sqrt{N}\Omega_0$. Additionally, the cooperative radiation pressure force is weighted by the difference of structure factors, $s_N-f_N$, where the $s_N$ part corresponds to the cooperative absorption process and the $f_N$ part to the cooperative emission. We will see that both forces may cancel completely in some cases.

\section{Structure factors for smooth density distributions}

Assuming a smooth Gaussian density distribution with ellipsoidal shape,
$n(\mathbf{r})=n_0\exp[-(x^2+y^2)/2\sigma_r^2-z^2/2\sigma_z^2]$, we can evaluate the structure factor by replacing the sum $\sum_j$ by an
integral $\int
d\mathbf{r}~n(\mathbf{r})$, and obtain
\begin{equation}\label{EqStructuregauss}
    S_\infty(k,\theta,\phi)=e^{-\sigma^2\left[\sin^2\theta+\eta^2\left(\cos\theta-k_0/k\right)^2\right]/2}~,
\end{equation}
where $\sigma=k\sigma_r$ and $\eta=\sigma_z/\sigma_r$ is the aspect ratio. Setting $k\simeq k_0$, one can show that for elongated clouds, $\eta\ge1$,
\begin{eqnarray}
    s_\infty^{(\eta)}  = \frac{\sqrt{\pi}e^{\frac{\sigma^2}{\eta^2-1}}}{4\sigma\sqrt{\eta^2-1}}\left\{\text{erf}\left[\frac{\sigma(2\eta^2-1)}
        {\sqrt{\eta^2-1}}\right]-\text{erf}\left[\frac{\sigma}{\sqrt{\eta^2-1}}\right]\right\}~,\nonumber\\
    f_\infty^{(\eta)}  = \frac{1}{\eta^2-1}\left[\eta^2 s_\infty^{(\eta)}-\frac{1}{4\sigma^2}(1-e^{-4\eta^2\sigma^2})\right]~.
\end{eqnarray}
For spherical clouds ($\eta=1$) one finds
\begin{eqnarray}\label{EqsNfNSmooth}
    s_\infty  =\frac{1}{4\sigma^2}\left[1-e^{-4\sigma^2}\right] \stackrel{\sigma\gg1}{\longrightarrow}  (2\sigma)^{-2}~,\\
    f_\infty  =\frac{1}{4\sigma^2}\left[1-\frac{1}{2\sigma^2}+\left(1+\frac{1}{2\sigma^2}\right)e^{-4\sigma^2}\right]
        \stackrel{\sigma\gg1}{\longrightarrow} s_\infty-2s_\infty^2~.\nonumber
\end{eqnarray}
For $\sigma,\eta\gg1$, $s_\infty^{(\eta)}$ can be approximated by
$s_\infty^{(\eta)}\simeq s_\infty\sqrt{\pi}Fe^{F^2}
\left[1-\text{erf}(F)\right]$, where
$F\equiv\sigma/\eta=k\sigma_r^2/\sigma_z$ is the Fresnel number.
For large Fresnel numbers $s_\infty^{(\eta)}\rightarrow s_\infty$.
    \begin{figure}[ht]
        \centerline{\scalebox{.5}{\includegraphics{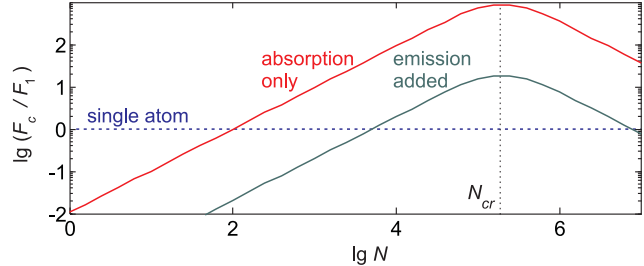}}}
        \caption{(Color online) The red solid line shows the $N$-dependence of the radiation pressure force resulting from pump photon absorption
        only, for $\eta=1$, $\sigma=5$ and $\Delta_0=10^3 \, \Gamma$.
            The green solid line traces the radiation pressure force, when photon emission is taken into account, and the blue dashed line traces the
            single atom radiation pressure.}
        \label{Fig2}
    \end{figure}

\section{Discussion} For a smooth-density spherical cloud the ratio between the cooperative radiation pressure
force (\ref{EqCooperativeforce}) and the single-atom force becomes
\begin{equation}\label{EqRatio}
    \frac{F_c}{F_1}=\frac{4\Delta_0^2+\Gamma^2}{4\Delta_0^2+N^2\Gamma^2s_\infty^2}~N s_\infty\left(1-\frac{f_\infty}{s_\infty}\right)~.
\end{equation}
The emission pattern is described by the $f_\infty$ term. An isotropic emission pattern, obtained e.g.~for small volumes, leads to
vanishing $f_\infty$. In contrast, for large volumes the recoil at emission compensates the recoil at absorption, $f_\infty\approx~s_\infty$,
which results in mainly forward emission. The above expression also shows that even for isotropic emission pattern, cooperative effects can
strongly modify the radiation pressure by a modified cooperative absorption (see Fig.~\ref{Fig2}). The dependency of Eq.~(\ref{EqRatio}) on
atom number exhibits a maximum at $N_{cr}=2\Delta_0/\Gamma s_\infty=4\sigma^2(2\Delta_0/\Gamma)$. Assuming large detuning $\Delta_0\gg\Gamma$,
the modification of the radiation pressure force for small atom numbers, $N\ll N_{cr}$, is
\begin{equation}
    \frac{F_c}{F_1}\approx\frac{2N}{(2\sigma)^4}~.
\end{equation}
On the other hand, for large atom numbers, $N\gg N_{cr}$, the radiation pressure becomes independent of the cloud size,
\begin{equation}
    \frac{F_c}{F_1}\approx\frac{2}{N}\left(\frac{2\Delta_0}{\Gamma}\right)^2~.
\end{equation}
The maximum radiation pressure, obtained at $N=N_{cr}$, is $F_c/F_1\approx\Delta_0/2\Gamma\sigma^2$ and hence, for large volumes with
$\sigma>\sqrt{\Delta_0/2\Gamma}$, below the radiation pressure expected for uncorrelated scattering. For small volumes the cooperative can exceed the uncorrelated radiation pressure.

It is interesting to compare this with the Dicke limit of small clouds, $|\mathbf{r}_j|\ll\lambda$, where the structure factor becomes $S_N(\mathbf{k})\simeq 1$. Consequently, the surface-integrated values are $s_N=1$, $f_N=0$ and $N_{sr}=2\Delta_0/\Gamma$, so that the force ratio for small and large atom numbers becomes, respectively,
\begin{equation}
    \frac{F_c}{F_1}\approx N \text{\ \ \ \ \ \text{and} \ \ \ \ \ } \frac{F_c}{F_1}\approx\frac{1}{N}\left(\frac{2\Delta_0}{\Gamma}\right)^2~.
\end{equation}

Note that the phase shift of the laser beam after passing through the atomic cloud can affect the driving term of Eq.~(\ref{EqHamiltonian}). Indeed, for a spherical atomic cloud with resonant optical density $b_0=(3\lambda^2/2\pi)\int dz~n(0,0,z)=3N/\sigma^2$, the characteristic atom number $N_{cr}$ can be expressed in terms of the phase shift experienced by the pump laser beam on its path across the cloud: $\Delta\phi=b_0\Gamma/(4\Delta_0) =6N/N_{cr}$. In other words, for atom numbers so large, that despite its large detuning the pump laser beam is shifted by more than $2\pi$, the radiation pressure force is dramatically suppressed. For low spatial densities of the atomic cloud, this phase shift remains small compared to the free propagation phase shift $k\sigma_z$ and only weakly modifies the scattering properties of the atomic cloud. As we assume that each atom is only driven by the external field, we also neglect multiple scattering of a photon in the atomic cloud. The above calculations are thus not valid for $b_0>4\Delta_0^2/\Gamma^2$.

The collective radiation pressure described in this paper is likely to be of importance for samples with large on-resonance optical thickness ($b_0\gg~1$). Samples with such
an optical thickness are e.g. used to
study multiple scattering of light in cold atoms \cite{Labeyrie2003} (with $b_0\approx40$) and most Bose-Einstein condensates also fulfill this criterium.
For an experimental observation of the collective radiation pressure, a low intensity quasi-resonant laser would be required to measure radiation pressure, e.g. in a time of flight experiment after the exchange of several photons per atom on average. Such experiments are state of the art as well for cold atom in magneto-optical traps as for atoms from a Bose-Einstein condensate.
For Bose-Einstein condensates the modification of the scattering properties of light might also allow to address the problem of spurious heating in optical dipole traps where spontaneous scattering might appear in some limits.

We note again that cooperativity in the radiation pressure as described in this paper does not rely on self-organization or correlations between different scattering events (as in \cite{Vuletic00, Deutsch1994}) but it is intrinsic in the scattering of a single photon by a correlated system of  $N$ atoms.
Although the calculated expression of the average radiation pressure does not exhibit specific quantum features and could be in principle obtainable classically, the study of higher moments of the atomic distribution and fluctuations should shed light on the quantum properties of the entangled state \cite{Eberly06, vanEnk2005}. This topic will be at the aim of our forthcoming studies.

\section{Conclusion} We have shown by an explicit calculation that the radiation pressure depends on the atom number via cooperative scattering. The assumption of a smooth density distribution allowed us to integrate the structure factors. We have shown that the radiation pressure is strongly affected by both a cooperative modification of the absorption and the emission properties of the atomic cloud. We have identified different regimes, where enhanced absorption can increase the radiation pressure for intermediate atom numbers or where forward emission almost cancels the recoil imparted to the atoms at the absorption. These calculations indicate that experiments are in reach with state of the art clouds of cold atoms. As the driven timed Dicke states involved in our model are entangled atomic states, radiation pressure might become an interesting tool to investigate non classical features of cooperative scattering of light by clouds of cold atoms.

\acknowledgments{This work has been supported by ANR CAROL (project ANR-06-BLAN-0096) and by the Deutsche Forschungsgemeinschaft (DFG) under Contract No. Co~229/3-1. S.B. acknowledges a grant from the Deutscher Akademischer Austauschdienst and E.L. support from INTERCAN.}

\end{document}